\documentclass[12pt]{article}

\usepackage{amsmath}
\usepackage{graphicx,psfrag,epsf}
\usepackage{enumerate}
\usepackage{natbib}
\usepackage{url} % not crucial - just used below for the URL

\newcommand{\blind}{0}

\usepackage{geometry}
\usepackage[utf8]{inputenc}

\usepackage{amsfonts,amssymb,bm}
\usepackage{mathptmx}%{times}
\usepackage{newtxtext}
\usepackage{newtxmath}

\newcommand{\bx}{\mathbf{x}}

\usepackage{xcolor}

\newcommand{\red}[1]{\textcolor{red}{#1}}

\usepackage{graphicx}
\usepackage{float}

\usepackage{algorithm}
\usepackage{algorithmic}
\floatname{algorithm}{Procedure}

\usepackage{verbatim}

\newcommand{\dc}[1]{#1^{ \circ}\mathrm{C}}

\newcommand{\bs}{\mathbf{s}}
\newcommand{\qtst}{q(\tau | \bs, t)}
\usepackage[caption = false]{subfig}
\usepackage[export]{adjustbox}
\usepackage{longtable}

\begin{document}
%\linenumbers
\def\spacingset#1{\renewcommand{\baselinestretch}%
	{#1}\small\normalsize} \spacingset{1}

%%%%%%%%%%%%%%%%%%%%%%%%%%%%%%%%%%%%%%%%%%%%%%%%%%%%%%%%%%%%%%%%%%%%%%%%%%%%%%

\if0\blind
{
	\title{\bf {Spatio-temporal quantile regression analysis revealing more nuanced patterns of climate change: a study of long-term daily temperature in Australia}}
	\author{Qibin Duan, Clare A. McGrory,  Glenn Brown, Kerrie Mengersen and
		You-Gan Wang\thanks{
			corresponding author, email:you-gan.wang@qut.edu.au}\hspace{.2cm}\\ \\
		School of Mathematical Sciences, Queensland University of Technology, Australia,
 \\ARC Centre of Excellence for Mathematical \& Statistical Frontiers (ACEMS)}
	\maketitle
} \fi

\bigskip

\begin{abstract}
Climate change is associated with an overall increase in mean global temperatures in recent years. Many studies consider temperature trends at the global scale, but the literature is lacking in in-depth analysis of the temperature trends across Australia in recent decades.
Australia comprises a very wide variety of environmental systems within the same country. It is therefore of great interest to analyse the spatio-temporal patterns of temperature across Australia. Modeling of daily Australia temperature data suffers from a quasi-periodic heterogeneity issue in variance that make analysis less reliable, but such an issue has not been overlooked in climate research. A contribution of this study is that we propose a joint model of quantile regression and variability in order to account for all heterogeneity of data.
A spatio-temporal version of the joint model is adapted to analyze the Australian daily maximum (Dmx) and minimum (Dmn) temperature data
(Jan 1, 1960 to Dec 31, 2019) so as to acquire more accurate spatio-temporal patterns.

According to the State of the Climate by \cite{BOMCSIRO2020}, Australia's climate has warmed by  $\dc{1.44\pm 0.24}$ since 1910, with most of the warming having occurred since 1950. As is reported in more detail in the Climate Statement by \citet{BOM2019}, the overall mean (average of daily maximum and minimum) temperature for the 10 year period from 2010 to 2019 was the highest on record, at $\dc{0.86}$ above the long term overall average since 1910, and $\dc{0.31}$ warmer than the 10 years 2000–2009, which is the second warmest 10-year period. Moreover, the climate change within Australia is also spatially varied. For example, for the wet tropic area in far North Queensland, the mean temperature increased by $\dc{1.1}$ between 1910 and 2013, while in the east coast area in New South Wales, the mean temperature increased by $\dc{0.8}$ in the same period. Also, for the Rangelands area, the north part increased $\dc{1}$ and the south part increased $\dc{0.9}$ during the same period. These statistical analyses undertaken in climate reports in Australia are based on inference about the maximum, minimum or mean temperature, and did not account for spatio-temporal correlations and quasi-periodic heterogeneity in variance.   

Our quantile regression is to estimate the conditional quantiles of daily temperature and can model all quantile levels and spatio-temporal correlations jointly with consideration of heterogeneity. This enables us to characterize the entire density function of daily maximum/minimum temperatures as a function of time for each location.
  Interestingly, our results suggest different patterns of climate change for different percentiles of daily maximum and minimum temperature series over Australia spatially.
Overall, the country appears to be experiencing a warming in daily temperature, where Dmx is warming $\sim\dc{0.08}$ per decade more than Dmn. For Dmx, the average warming trend of all 72 stations is 0.22, 0.22 and $\dc{0.20}$ per decade for quantile levels 0.1, 0.5 and 0.9, receptively, while these values become 0.14, 0.13, $\dc{0.14}$ for Dmn. In Dmx series, the lower quantiles generally increase more than the higher quanitles.

In more details of Dmx series, for 0.1 quantile, the trend of all 72 stations are warming. The north-east area including NSW, VIC, TAS, south part of QLD and east part of SA appear to experience more of an increase than other regions under 0.1 quantile of Dmx, with $\geq \dc{0.3}$ increase per decade in the last 60 years. This area covers the most part of the Murray–Darling basin, which experienced water loss in the last few years. We especially highlight TAS and greater Sydney area which increase the most with nearly $\dc{0.4}$.  In contrast, the far north QLD and north-west corner of WA have least change with around $\dc{0.1}$ increase in the same period.
In terms of the 0.5 quantile of Dmx, TAS and parts of NSW still have mostly increasing temperatures with  $>\dc{0.3}$ per decade, while other areas have $\dc{0.1}$ to $\dc{0.3}$ increase, except for the north of QLD and north-west corner of WA showing a non-significant increase $\leq\dc{0.1}$. When it comes to the 0.9 quantile, parts of the north of QLD show a non-significant increase, or even decrease. Other regions generally have a warming trend of $\dc{0.2}$ to $\dc{0.3}$.

The trend of Dmn series perform differently. For 0.1 quantile of Dmn,  south QLD, small region of north coast of WA, coastal area near Sydney and Adelaide show a warming trend around $\geq\dc{0.3}$. These areas continue to show warming trend for the 0.5 quantile, but not for the 0.9 quantile. While other regions have less increase with $\leq\dc{0.2}$ for 0.1 quantile, especially some areas (south-east and south-west region of Australia) show a cooling trend with negative value of trend function.
For 0.5 quantile, most part of Australia does not show a significant increase, with most values of trend function lying 0 to $\dc{0.1}$, while the south-east area experiences a consistent cooling. However, no significant cooling trend is showed for the 0.9 quantile. Instead, inland part of QLD and north coast of WA show an increase with $~\dc{0.3}$.

\end{abstract}

\vfill

\newpage

\spacingset{1.45} % DON'T change the spacing!

\section{Introduction}
The increase in the intensity and frequency of global extreme weather events, e.g., extreme heat, extreme cold, drought, snow cover decline, is attributed to  fundamental changes in the underlying climate  \citep[see][]{JC2,JC1}. Unchecked, climate change, is hence a major threat now faced by the entire world because of the severe societal and ecological impacts it will have. There is an overall warming trend occurring with the global mean temperature estimated to have risen by  $~\dc{0.85}$ during 1880-2012 \citep{IPCC}.
This increasing trend is predicted to continue, and if it does, auguring huge environmental and social change.
Human activity is widely argued to be the major cause of  global warming \citep{IPCC2007,IPCC2014}, but a great deal of uncertaintly still remains regarding the exact mechanism which underlies the warming process \citep{Sinha}. As \cite{Sinha} highlights, the global surface warming of recent decades has been realized as a succession of periods of warming slowdowns, or hiatus, followed by warming surges.

 While warming is the trend globally, at the regional and local scales, a wide variety of changes in temperature are observable across the globe \citep{Alexander2006,Brown2008}. Climate changes have been extensively studied at the global scale  \citep[see][]{Zhang}), but within Australia itself, in-depth statistical studies on historic temperature changes are lacking \citep{QldTempRain}.  Australia is the largest country in Oceania and one of the major producers of agricultural products, so deep understanding of climate change in Australia is of importance.

According to the State of the Climate \cite{BOMCSIRO2018}, Australia's climate has warmed by just over $\dc{1}$ since 1910,  which is specified as $\dc{1.44\pm 0.24}$ in \cite{BOMCSIRO2020}, with most of the warming having occurred since 1950.  This has been accompanied by increased frequency of extreme heat events.
As is reported in more detail in the Climate Statement by \citet{BOM2019}, the overall mean (average of daily maximum and minimum) temperature for the 10 year period from 2010 to 2019 was the highest on record, at $\dc{0.86}$ above the long term overall average since 1910, and $\dc{0.31}$ warmer than the 10 years 2000–2009, which is the second warmest 10-year period. The climate change within Australia is also spatially varied. For example, for the wet tropic area in far North Queensland, the mean temperature increased by $\dc{1.1}$ between 1910 and 2013, while in the east coast area in New South Wales, the mean temperature increased by $\dc{0.8}$ in the same period. Also, for the Rangelands area, the north part increased $\dc{1}$ and the south part increased $\dc{0.9}$ during the same period.

The majority of statistical analyses undertaken in climate studies in Australia are based on inference about the maximum, minimum or mean temperature. In this study we take an approach based on quantile regression estimates of the data with consideration of spatial correlation. Quantile regression was originally proposed by \citet{Koenker1978} as an alternative approach to mean regression that does not require the usual strict assumption about normally distributed residuals in the regression models. The idea has been used to identify changes over time of any percentiles of climate variables;  see \citet{koenker1994quantile,Barbosa2011,gao2017quantile,Franzke2013} for particular focus on the analysis of temperature series. {Quantile regression is particularly useful, as Australia (possibly also other countries) daily temperature series are merely normally distributed}. The performance of quantile regression regarding to trend detection analysis has been compared with traditional approaches, i.e.,robust linear regression and the nonparametric M-K test, showing that quantile regression is reliable \citep{gao2017quantile}. The spatial and temporal variation can be modeled jointly as  in \citet{Brian2013}, in a Bayesian manner.

  One advantage of this approach is that quantile regression estimates are less influenced by extreme outliers in the response measurements than standard linear regression estimates based on the mean are. The other big motivation for this approach is that the conditional quantile functions are of interest to us. {The variety of measures} of central tendency and statistical dispersion allow us to describe the relationship at different points in the conditional distribution of the outcome. In this way we obtain a more comprehensive picture of the relationship between the variables. When using quantile regression, we have increased freedom in our modelling in that we are not restricted by the assumption that variable relationships are the same at the median and tails of the distribution as they are at the mean. {Such models could provide a more} in-depth understanding of the historic change in Australian temperature, which can potentially improve anticipation and management of climate risk and associated negative impacts.

We highlight however,  that in analysis of daily Australia temperature data, complicated heterogeneity in variance arises and this must be addressed if results are to be reliable.
	In this article we propose a joint model including variance as a covariate in our spatio-temporal regression to do so. We show through simulation studies that this is a sensible approach to dealing with the issue. The proposed model allows us to characterize the entire quantile process over time with accommodation of trend, asymmetry, heterogeneity and seasonality.  

The remainder of this report is organized as follows. The data and study region are described in Section 2.  Section 3 presents the results of our real data analysis and Section 4 concludes the study.
The exploratory data analysis and methods of formal data analysis with a simulation study are described in Appendix. 
\section{Sources of data}
\subsection{Study Region}
In this study we analyze the spatio-temporal pattern of warming in Australia based on the trends of averages and quantiles of daily maxima and minima temperatures using recent observed data.

The geographical boundary of Australia is between $9^{\circ}- 44^{\circ}S$ latitude and  $112^{\circ} - 154^{\circ}E$ longitude (apart from Macquarie Island), with total area of $7,692,024$ $\mathrm{km}^2$. The massive size of the country gives it a wide variety of landscapes, with tropical rainforests in the north-east, mountain ranges in the south-east, south-west and east, and desert and semi-arid land in the center, resulting in a variety of climates across the country.
%The climate in Australia is also significantly influenced by ocean currents, including the Indian Ocean Dipole and the El Niño–Southern Oscillation, and the seasonal tropical low-pressure system that produces cyclones in northern Australia \red{[We need references for this]}.
The northern part of the country has a tropical climate, with predominantly summer rainfall. As for the southern part, the south-west corner of the country has a Mediterranean climate, while the south-east ranges from oceanic (Tasmania and coastal Victoria) to humid subtropical (from the upper half of New South Wales), with highlands featuring alpine and subpolar oceanic climates. The interior desert has stable arid and semi-arid climates. {More details about Australian climate can be found in the website of \cite{ClimateInformation}.}

Considering the real world impacts of variations in Australian temperatures in more detail, we give some examples of areas will effects will be most pronounced:
\begin{itemize}
	\item  A heatwave occurs when the daily maximum and minimum temperatures are unusually hot over a three-day period at a location. Heat waves affect human health, transportation, electricity demand and infrastructure performance, outside labor productivity, and ecology and are a risk to natural systems (e.g., bush fires).
	\item  Duration of elevated temperatures affecting human health, outside labour productivity, transport infrastructure, electricity demand etc. For instance, CSIRO research shows that 4 sequential days of elevated temperature at $\dc{35}$ requires 33\% more cooling energy for a building than does 4 distributed days at the same temperature.
	\item  The daily minimum temperatures affect incidence of frosts required for key processes for sweetening fruit, e.g., citrus, grapes etc.
	\item  Daily and seasonal temperature range affects agricultural productivity in terms of crop yields and protein content, transportation.
	\item Rate of change of warming affecting global risk by reducing the time available to adapt to changing conditions
\end{itemize}

\subsection{Data collection and selection}

The daily maxima (Dmx) and minima (Dmn) temperature series were obtained from 1,745 weather stations operated by the Bureau of Meteorology (BoM) Australia. These datasets can be accessed through the \texttt{R} package \texttt{bomrang} \citep{bomrang}. The length of these time series varies among different stations from less than 10 years to more than 100 years of observations. The latitude, longitude and elevations of each station are also obtained from the BoM website.

As the years of 2000 and 2010 are the starting years of the top two warmest 10-year periods in Australia on record, we considered data from stations not covering these periods to be less meaningful in terms of studying the warming trend in Australia. Therefore, we excluded stations that were not operating during these periods.

\begin{figure}[H]
	\centering
	\includegraphics[width=0.8\textwidth]{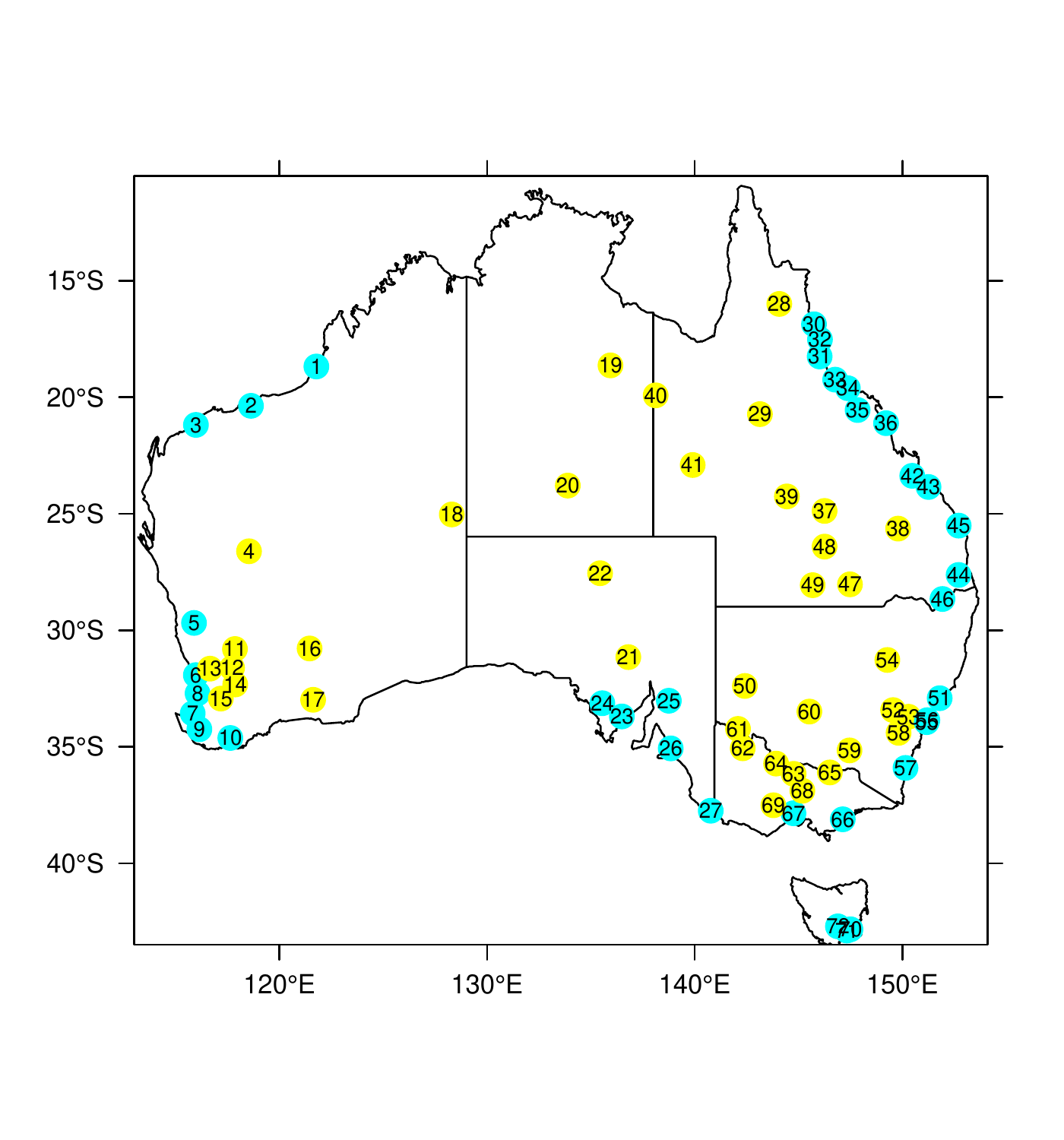}
	\caption{The distribution of selected stations with qualifying data. Blue dots are for (near) coastal stations and yellow dots are for inland stations.}
	\label{fig.distribution}
\end{figure}

{We restrict the study period from Jan 1, 1960 to Dec 31, 2019, so that most warming periods can be covered \citep{gmd-13-5175-2020} and more operating stations can be included}.  Furthermore, we exclude the stations with over $20\%$ of missing observations.  As a result, data from 72 stations is included in our analysis as shown in TABLE~\ref{tab.stations}. Specifically, there are 22 stations in Queensland, 18 in Western Australia, 11 in New south Wales, 9 in Victoria, 7 South Australia, 4 in Tasmania and 2 in the Northern Territory.  They are geographically distributed as shown in FIG~\ref{fig.distribution}, and the ID number of each station is also accompanied for the sake of description of result.

\renewcommand\arraystretch{0.6}
\begin{table}[H] \small
	
	\caption{Distribution of selected stations and their details}\label{tab.stations}
	\begin{longtable}{c | c | c c| c || c |c | c c| c}
Station ID	&	BoM Station ID	&	Lat	&	Lon	&	State	&	Station ID	&	BoM Station ID	&	Lat	&	Lon	&	State	\\ \hline
1	&	3030	&	-18.7	&	121.8	&	WA	&	37	&	35069	&	-24.9	&	146.3	&	QLD	\\
2	&	4032	&	-20.4	&	118.6	&	WA	&	38	&	35070	&	-25.6	&	149.8	&	QLD	\\
3	&	5008	&	-21.2	&	116.0	&	WA	&	39	&	36026	&	-24.3	&	144.4	&	QLD	\\
4	&	7045	&	-26.6	&	118.5	&	WA	&	40	&	37010	&	-19.9	&	138.1	&	QLD	\\
5	&	8025	&	-29.7	&	115.9	&	WA	&	41	&	38003	&	-22.9	&	139.9	&	QLD	\\
6	&	9021	&	-31.9	&	116.0	&	WA	&	42	&	39083	&	-23.4	&	150.5	&	QLD	\\
7	&	9534	&	-33.6	&	115.8	&	WA	&	43	&	39123	&	-23.9	&	151.3	&	QLD	\\
8	&	9538	&	-32.7	&	116.1	&	WA	&	44	&	40004	&	-27.6	&	152.7	&	QLD	\\
9	&	9573	&	-34.3	&	116.1	&	WA	&	45	&	40126	&	-25.5	&	152.7	&	QLD	\\
10	&	9581	&	-34.6	&	117.6	&	WA	&	46	&	41095	&	-28.7	&	151.9	&	QLD	\\
11	&	10007	&	-30.8	&	117.9	&	WA	&	47	&	44010	&	-28.0	&	147.5	&	QLD	\\
12	&	10073	&	-31.6	&	117.7	&	WA	&	48	&	44021	&	-26.4	&	146.3	&	QLD	\\
13	&	10111	&	-31.7	&	116.7	&	WA	&	49	&	44026	&	-28.1	&	145.7	&	QLD	\\
14	&	10536	&	-32.3	&	117.9	&	WA	&	50	&	47019	&	-32.4	&	142.4	&	NSW	\\
15	&	10614	&	-32.9	&	117.2	&	WA	&	51	&	61055	&	-32.9	&	151.8	&	NSW	\\
16	&	12038	&	-30.8	&	121.5	&	WA	&	52	&	63005	&	-33.4	&	149.6	&	NSW	\\
17	&	12071	&	-33.0	&	121.6	&	WA	&	53	&	63039	&	-33.7	&	150.3	&	NSW	\\
18	&	13017	&	-25.0	&	128.3	&	WA	&	54	&	64008	&	-31.3	&	149.3	&	NSW	\\
19	&	15085	&	-18.6	&	135.9	&	NT	&	55	&	66037	&	-33.9	&	151.2	&	NSW	\\
20	&	15590	&	-23.8	&	133.9	&	NT	&	56	&	66062	&	-33.9	&	151.2	&	NSW	\\
21	&	16001	&	-31.2	&	136.8	&	SA	&	57	&	69018	&	-35.9	&	150.2	&	NSW	\\
22	&	17043	&	-27.6	&	135.4	&	SA	&	58	&	70080	&	-34.4	&	149.8	&	NSW	\\
23	&	18014	&	-33.7	&	136.5	&	SA	&	59	&	72150	&	-35.2	&	147.5	&	NSW	\\
24	&	18044	&	-33.1	&	135.6	&	SA	&	60	&	75032	&	-33.5	&	145.5	&	NSW	\\
25	&	19062	&	-33.0	&	138.8	&	SA	&	61	&	76031	&	-34.2	&	142.1	&	VIC	\\
26	&	23733	&	-35.1	&	138.8	&	SA	&	62	&	76047	&	-35.1	&	142.3	&	VIC	\\
27	&	26021	&	-37.7	&	140.8	&	SA	&	63	&	80015	&	-36.2	&	144.8	&	VIC	\\
28	&	28004	&	-16.0	&	144.1	&	QLD	&	64	&	80023	&	-35.7	&	143.9	&	VIC	\\
29	&	30045	&	-20.7	&	143.1	&	QLD	&	65	&	82039	&	-36.1	&	146.5	&	VIC	\\
30	&	31011	&	-16.9	&	145.7	&	QLD	&	66	&	85072	&	-38.1	&	147.1	&	VIC	\\
31	&	32004	&	-18.3	&	146.0	&	QLD	&	67	&	87031	&	-37.9	&	144.8	&	VIC	\\
32	&	32025	&	-17.5	&	146.0	&	QLD	&	68	&	88109	&	-36.9	&	145.2	&	VIC	\\
33	&	32040	&	-19.2	&	146.8	&	QLD	&	69	&	89002	&	-37.5	&	143.8	&	VIC	\\
34	&	33002	&	-19.6	&	147.4	&	QLD	&	70	&	94008	&	-42.8	&	147.5	&	TAS	\\
35	&	33013	&	-20.6	&	147.8	&	QLD	&	71	&	94029	&	-42.9	&	147.3	&	TAS	\\
36	&	33119	&	-21.1	&	149.2	&	QLD	&	72	&	95003	&	-42.7	&	146.9	&	TAS	\\		
	
	\end{longtable}
\end{table}

\section{Results}
%\subsection{Fitting results}
%FIG~\ref{fig.fitting} shows the fitting curves of daily maximum temperature for 6 stations.  \red{(Do we need some %criteria or justification for the model fitting? I try to replaced FIG 3 with a table to show the fitting. )}
%\begin{figure}[H]
%	\centering
%	\includegraphics[width=0.8\textwidth]{fig_qrfitting_all10.pdf}
%	\caption{Fitting of quantile regression for 6 stations with daily maximum data.}
%	\label{fig.fitting}
%\end{figure}

\subsection{Quantile trend analysis}
\begin{figure}[H]
\includegraphics[width=1.1\textwidth]{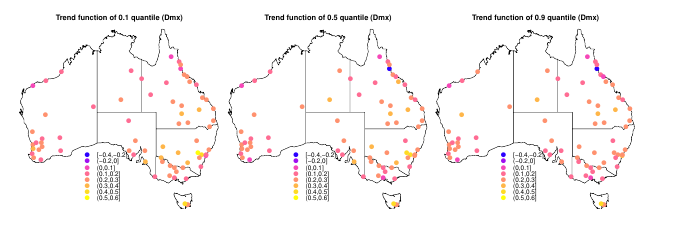}
\caption{Quantile trend of Dmx for all 72 stations during the study period. The color of point is the total degrees Celsius that increased/decreased per decade from 1960 to 2019.}
\label{fig.mapdots.max}
\end{figure}

\begin{figure}[H]
	\includegraphics[width=1.05\textwidth]{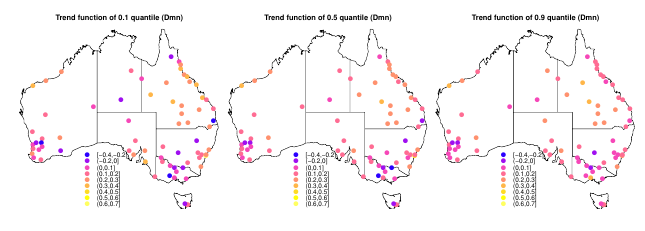}
	\caption{Quantile trend of Dmn for all 72 stations during the study period. The color of point is the total degrees Celsius that increased/decreased per decade from 1960 to 2019.}
	\label{fig.mapdots_min}
\end{figure}

\begin{figure}[H]
	\includegraphics[width=0.9\textwidth]{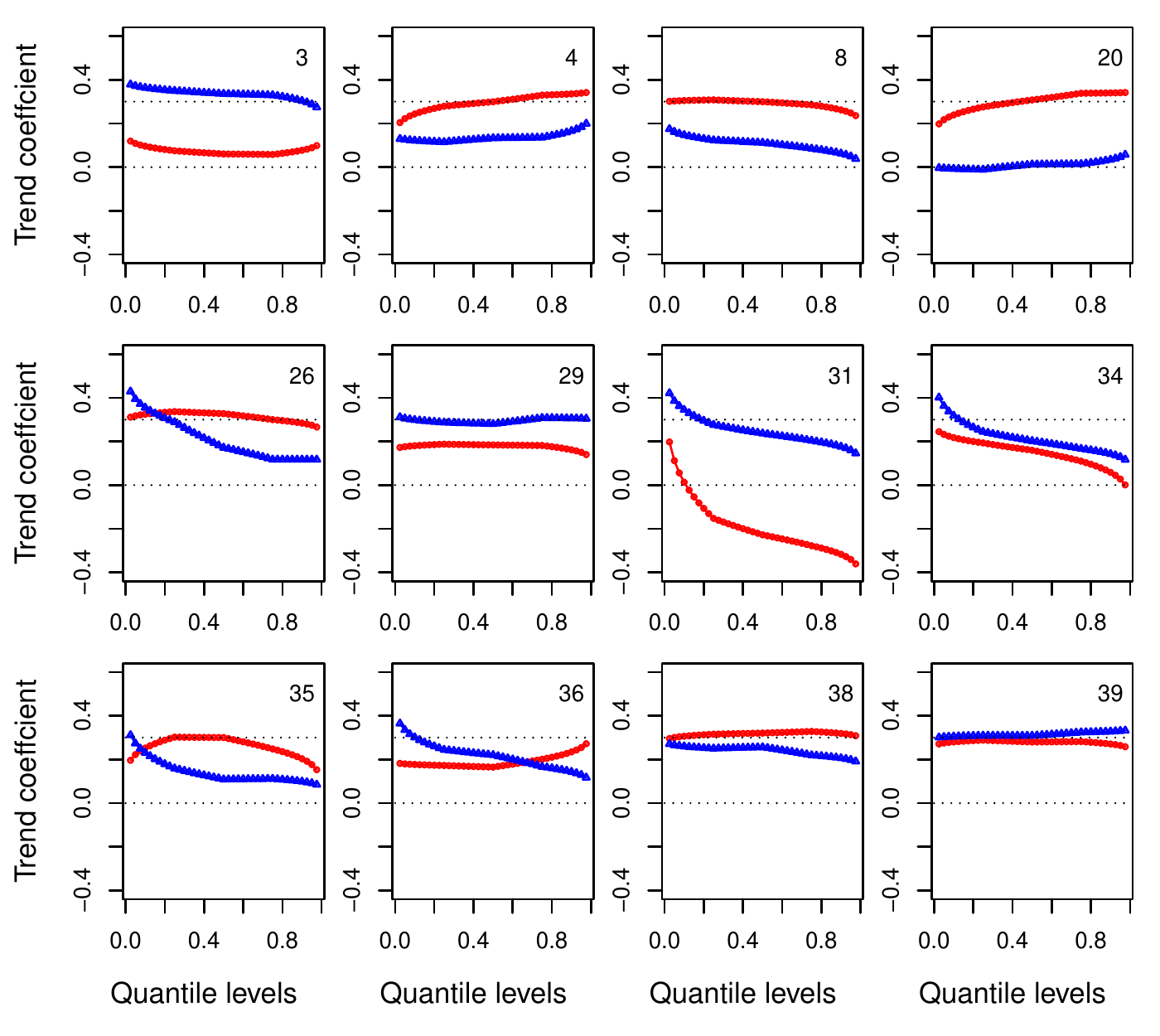}
	\caption{Quantile trend as a function of quantile level for selected stations. The red curve is for Dmx and the blue curve for Dmn. }
	
	\label{fig.qt.mxmn1}
\end{figure}

\begin{figure}[H]
	\includegraphics[width=0.9\textwidth]{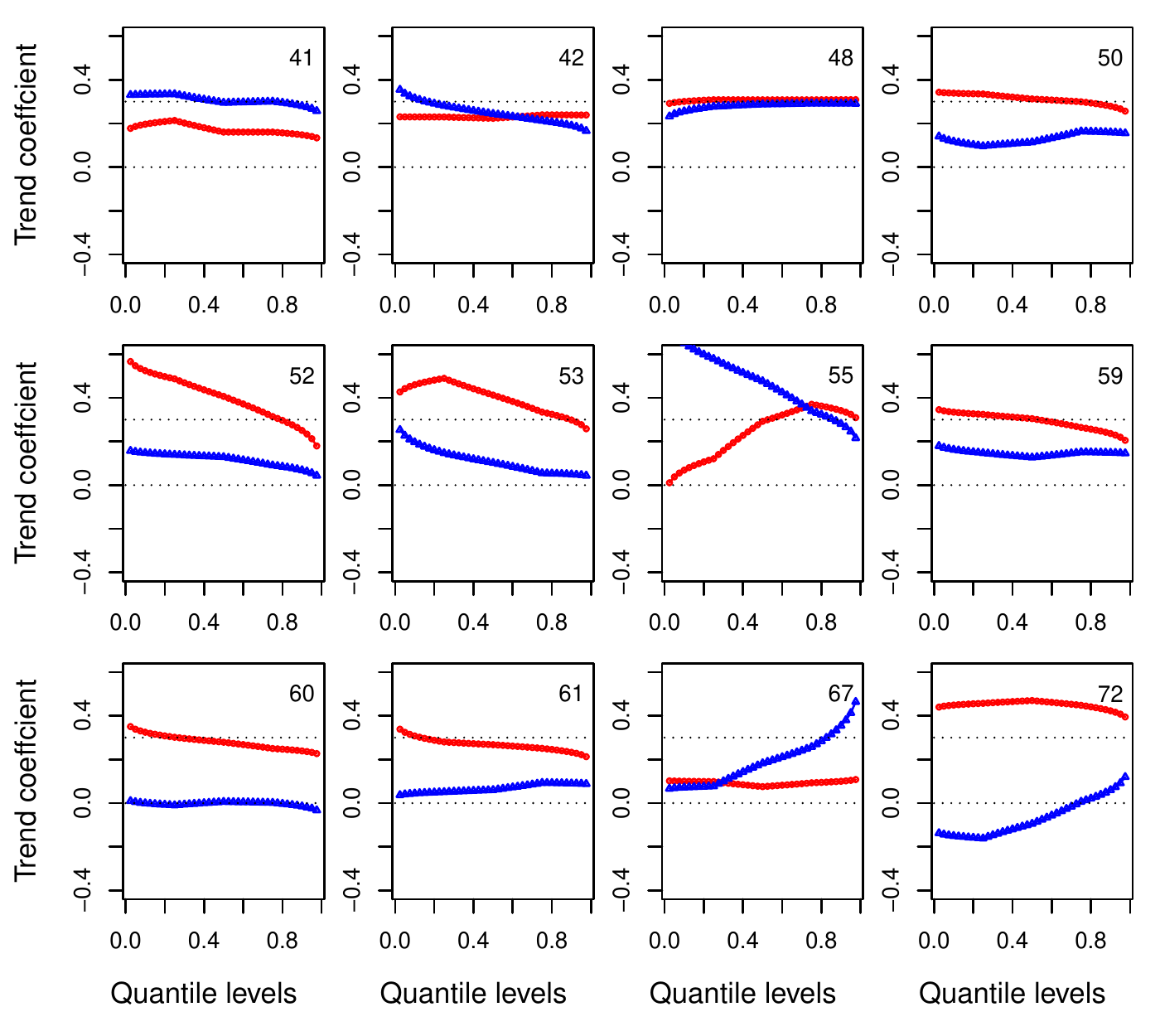}
	\caption{Quantile trend as a function of quantile level for selected stations. The red curve is for Dmx and the blue curve for Dmn.}
	
	\label{fig.qt.mxmn2}
\end{figure}

FIG~\ref{fig.mapdots.max} and \ref{fig.mapdots_min} show the values of trend function of quantiles 0.1, 0.5 and 0.9 for each of the included stations, respectively, while FIG~\ref{fig.qt.mxmn1} and \ref{fig.qt.mxmn2} show change of trend function with quantile levels for stations with any trend function $>\dc{0.3}$ . Note that the value represents the total increases in degrees Celsius per decade during the study period of 60 years. For daily maximum temperature, the average warming trend of all 72 stations is 0.22, 0.22 and $\dc{0.20}$ per decade for quantile levels 0.1, 0.5 and 0.9, receptively, while these values become 0.14, 0.13, $\dc{0.14}$ for daily minimum temperature. For overall trend, the changing rates of all three quantile levels have similar behaviors in both Dmx and Dmn, without presenting much heterogeneity. In comparison with the warming rate (range $\dc{0.12-0.16}$ per decade) of mean temperature since 1910 in BOM and CSIRO reports, the median of Dmx is warming with a bigger rate than reported value and the median of Dmn has a similar rate. When the period is reduced to post-1950 when the most of warming occurs, the warming rate of mean temperature (not reported by BOM and CSIRO) is bigger than that of pre-1950 period and the value is very likely to be within the warming rate of median of Dmn and Dmx (range $\dc{0.13-0.22}$ per decade).

For all three quantile levels, the daily maximum tempertures are increasing in all stations except for one in north Queensland (station 31 in FIG~\ref{fig.distribution} located in Cardwell Marine Pde QLD), and the values of trend function per decade mostly lie in $(0.1,0.3)$ indicating a warming trend of $\dc{0.6}$ to $\dc{1.8}$ in total during the last 60 years.
For quantile $0.1$ representing cold daily maximum temperture, there are five stations in NSW that increase by more than $\dc{0.3}$ per decade during the study period.
Three of which (stations 50, 59 60) are just within $(0.3,0.4)$ and decrease slightly to $(0.2,0.3)$ for higher quantiles, and another two (station 52, 53) of which are $>0.4$ and also decrease for higher quantiles but remain $>0.2$ in FIG~\ref{fig.qt.mxmn2}.  Also, other stations (station 38 and 48 in QLD, 8 in WA, 26 in SA, 72 in TAS) have been warming  $\geq\dc{0.3}$ for all three quantile levels per decade, especially for station 8 in TAS around $ \dc{0.4}$. And station 61 in VIC has been warming $>\dc{0.3}$ for 0.1 quantile, but this decreases to  $<0.3$ for higher quantiles.
For quantile 0.5 representing median daily maximum temperture, there are eight stations having $>\dc{0.3}$ increase per decade, in which station 20 in NT and 35 in QLD have lower increase in 0.1 quantile $<\dc{0.3}$ as shown in FIG~\ref{fig.qt.mxmn1}. And three have $>\dc{0.4}$ increase with two in NSW (station 52 and 53) and one (station 72) in TAS.
 In terms of quantile 0.9 for hot daily maximum temperture, six stations show the $>\dc{0.3}$ increase with two in QLD (station 38 and 48), and one in NSW (station 55), TAS (station 72), WA (station 4) and NT (station 20). The station 31 in QLD (located in Cardwell Marine Pde) shows declining trend under both 0.5 and 0.9 quantiles as indicated in FIG\ref{fig.mapdots.max}.
%
%\begin{figure}[H]
%	\includegraphics[width=0.5\textwidth]{figure2_dotsplot_01_tmax_de.pdf}
%	\includegraphics[width=0.5\textwidth]{figure2_dotsplot_05_tmax_de.pdf} \\ [-20pt]
%	\includegraphics[width=0.5\textwidth]{figure2_dotsplot_09_tmax_de.pdf} \\[-30pt]
%	\caption{Quantile trend of Dmx for all 72 stations for the study period. The color of point is the total degrees Celsius that increased/decreased per decade from 1960 to 2019.}
%	\label{fig.mapdots.max}
%\end{figure}
The pattern of the change in trend of Dmn is different from Dmx, as shown in FIG \ref{fig.mapdots_min}, where  the values of trend function per decade mostly lie in $[0.0, 0.3]$, and more stations have a cooling trend with negative values of trend functions.
The trend function of quantile 0.1 represents the change of temperature for the extremely cold days,  for which there are 12 stations that show a cooling trend and six of them have been cooling $>\dc{0.1}$ per decade during the study period, i.e., station 12 and 13 in WA, 44 in QLD, 65 and 69 in VIC, and 72 in TAS.
And nine stations show an increasing trend with amount $>\dc{0.3}$ per decade, i.e., station 3 in WA, station 26 in SA,  six stations in QLD (station 31,34,36,39,41 and 42), and station 55 in NSW ($>\dc{0.4}$).
For quantile level 0.5, eight stations show a cooling trend in the south east and south west of Australia with negative values of trend function, and four (station 12, 44, 65 and 69) have been cooling $>\dc{0.1}$ per decade.
Three stations (station 3, 39 and 55) in WA, NSW show a warming trend with amount $>\dc{0.3}$. For quantile level 0.9, only four stations still show a cooling trend with two (station 12 and 13) in WA, one (60) in NSW and one (63) in VIC. And three stations have an increasing trend with $>\dc{0.3}$ degrees with one (station 3) in WA, two (station 29 and 39) in QLD and one in NSW (station 67).

In terms of trend function against quantile levels, FIG~\ref{fig.qt.mxmn1} and \ref{fig.qt.mxmn2} display stations with any trend function $>\dc{0.3}$. Note that red curve above blue curve indicates that Dmx has increased more than Dmn, and vice versa.  If trend values increase with quantile level, it indicates bigger variation in the series, and vice versa. And a positive trend value in high quantile of Dmx means more extreme heat event and a negative value of trend function in low quantile of Dmn means more extreme cold events.
For example, stations 4, 8, 20, 35 and 38 in FIG~\ref{fig.qt.mxmn1} and stations 50, 52, 53, 59, 60, 61 and 72 in FIG~\ref{fig.qt.mxmn2} present such a pattern with red curve above blue curve. For station 4, higher quantiles increase more than lower quantiles for both Dmx and Dmn with all $>0$, indicating warming trend and bigger variation in both series, and more extreme heat events and less extreme cold events.  Station 8 has a pattern with warming trend but smaller variation in both series. While station 20 shows a warming trend in Dmx series with bigger variation and more extreme heat events. As for station 35, it presents a warming trend with more extreme heat events but no bigger variation in Dmx.  Station 52 and 53 present a similar pattern with trend coefficient decrease with quantile level in both series, which indicates a warming trend but smaller variation.  Station 60 and 61 have a warming trend in Dmx, but no clear trend in Dmn. Stations 50 and 59 have a warming trend in both series with smaller variation in Dmx.
Station 3, 29, 31, and 39  FIG~\ref{fig.qt.mxmn1} and  station 41 in FIG~\ref{fig.qt.mxmn2} have the blue curve above the red curve. Especially for station 31, the Dmx series present a cooling trend for trend coefficient of most quantile $<0$ and a smaller variation. The Dmn show a warming trend and a smaller variation.  Although there are clear trends in both Dmx and Dmn, when it comes to the mean temperature, such trend might be difficult to obtain. For this station, the extreme events occur less.

Note that some stations have red and blue curves crossed, e.g., station 36, 42, 55 and 67, where at that quanitle level both Dmx and Dmn have same trend.


\subsection{Spatial pattern of quantile trend}

\begin{figure}[H]

	\includegraphics[width=1.05\textwidth]{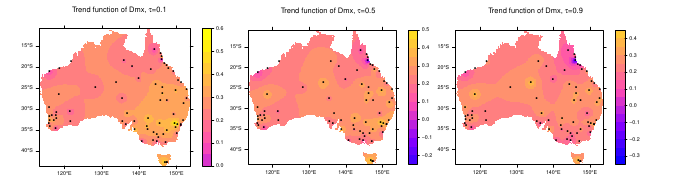}
	\caption{Quantile trend of Dmx for overall Australia during the study period. The color of point is the total degrees Celsius that increased/decreased per decade from 1960 to 2019.}
	\label{fig.mapsp1_tmax}
\end{figure}

\begin{figure}[H]
	\includegraphics[width=1.05\textwidth]{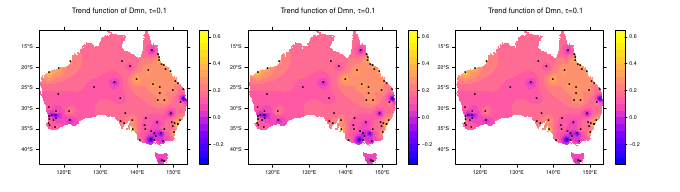}
	\caption{Quantile trend of Dmn for overall Australia during the study period. The color of point is the total degrees Celsius that increased/decreased per decade from 1960 to 2019.}
	\label{fig.mapsp1_tmin}
\end{figure}

The spatial pattern of trend functions across Australia is shown in Figure~\ref{fig.mapsp1_tmax} and \ref{fig.mapsp1_tmin} for the trend of Dmx and Dmn, respectively. For 0.1 quantile of Dmx, the values of trend function are all positive with warming trend for cold days (after removing seasonality). The north-east area including NSW, VIC, TAS, south part of QLD and east part of SA appear to experience more of an increase than other regions under  0.1 quantile of Dmx, with $\geq \dc{0.3}$ increase per decade in the last 60 years. This area covers the most part of the Murray–Darling basin, which experienced water loss in the last few years \citep[see][]{MDwaterloss}. We especially highlight TAS and greater Sydney area which increase the most with nearly $\dc{0.4}$.  In contrast, the far north QLD and north-west corner of WA have least change with around $\dc{0.1}$ increase in the same period.
In terms of the 0.5 quantile of Dmx, TAS and parts of NSW still have mostly increasing temperatures with  $>\dc{0.3}$, while other areas have $\dc{0.1}$ to $\dc{0.3}$ increase, except for the north of QLD and north-west corner of WA showing a non-significant increase $\leq\dc{0.1}$. When it comes to the 0.9 quantile, parts of the north of QLD show a non-significant increase, or even decrease. Other regions generally have a warming trend of $\dc{0.2}$ to $\dc{0.3}$.

\bigskip

The trend of Dmn series perform differently. For 0.1 quantile of Dmn,  south QLD, small region of north coast of WA, coastal area near Sydney and Adelaide show a warming trend around $\geq\dc{0.3}$. These areas continue to show warming trend for the 0.5 quantile, but not for the 0.9 quantile. While other regions have less increase with $\leq\dc{0.2}$ for 0.1 quantile, especially some areas (south-east and south-west region of Australia) show a cooling trend with negative value of trend function.
For 0.5 quantile, most part of Australia does not show a significant increase, with most values of trend function lying 0 to $\dc{0.1}$, while the south-east area experiences a consistent cooling. However, no significant cooling trend is showed for the 0.9 quantile. Instead, inland part of QLD and north coast of WA show an increase with $~\dc{0.3}$.

\subsection{Quantile trend by season}
\begin{figure}[H]
	\hspace{-0cm}	\includegraphics[width=1.05\textwidth]{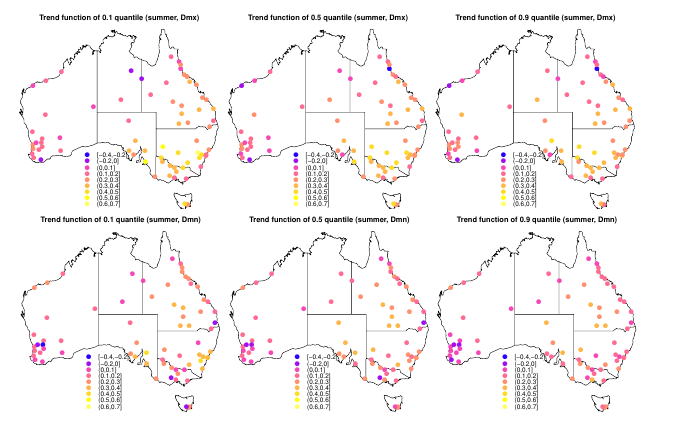}
		\caption{Quantile trend for $\tau = 0.1,0.5$ and $0.9$ for Dmx (top row) and Dmn (bottom) in summer (December, January and February) for all 72 stations. The color of point is the total degrees Celsius that increased/decreased per decade from 1960 to 2019.}
	\label{fig.mapdots.summer}
\end{figure}

This section shows the quantile trend for summer and winter seasons. Figure \ref{fig.mapdots.summer} shows the results for summer for three quantile 0.1, 0.5 and 0.9. Under all three quantiles, south-east part of Australia consistently have warming summer, including NSW, VIC, SA, TAS and south QLD, where most stations present a large increase $\geq \dc{0.3}$ per decade in last 60 years. In contrast,  the south-west region shows no significant trend of temperature for summer, with per decade increase $\leq \dc{0.1}$ for most stations in the region. The northern part of Australia generally has no significant trend of warming for most stations.

In terms of Dmn, stations in coastal area of NSW and inland area of QLD show an increase by $\geq\dc{0.2}$ for 0.1 quantile, and stations in inland area of QLD for 0.5 and 0.9 quantiles.  Generally in summer, daily maximum temperature (of both hot and cold days)  get much hotter in the south-east regions, while only slightly hotter in other regions. For daily minimum temperature, days get much warmer in south QLD in general. And cold days get much warmer in the NSW coastal region. Other regions experience a relatively smaller warming trend with most stations showing an increase $\leq\dc{0.2}$ and even decrease.

\begin{figure}[H]
	\hspace{-0cm}	\includegraphics[width=1.05\textwidth]{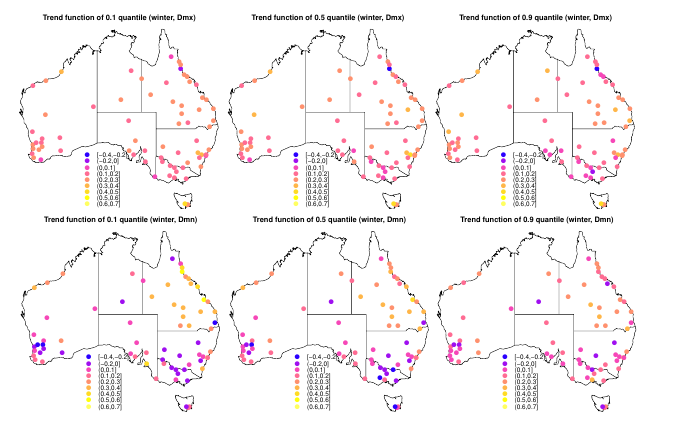}
	\caption{Quantile trend for $\tau = 0.1,0.5$ and $0.9$ for Dmx (top row) and Dmn (bottom) in winter (June, July and August) for all 72 stations. The color of point is the total degrees Celsius that increased/decreased per decade from 1960 to 2019}
	\label{fig.mapdots.winter}
\end{figure}

As for winter temperature in Figure~\ref{fig.mapdots.winter}, we can see from the Dmx series that the south part of Australia experiences less increase than the north part (QLD) except TAS having a station with substantial increase $\geq{0.4}$ under all three quantiles. Within QLD, the southern area shows an increasing trend $\geq \dc{0.2}$ that is more than the far north area where one station shows cooling trend.
The Dmn series also show that the northern area (especially QLD) increases more than southern region, but the difference is that the southern area (especially VIC, NSW, TAS and south part of WA) shows a clear cooling trend under all three quantiles, though quantile 0.9  decreases less than quantile 0.1 and 0.5. In QLD, generally higher quantiles increase less than lower quantiles, e.g., for quantile 0.1, most stations show an increase with $\geq \dc{0.3}$, while for quantile 0.9 the increase amounts are mostly in $\dc{0.2}$ to $\dc{0.3}$. One station in far north QLD shows non-significant increase or even decrease in these quantiles.

\section{Discussion}

This article has provided a detailed looked at the changes in temperature across Australia in recent decades.

In this study, we first raised the heterogeneity issue in the daily temperature time series with an exploratory analysis, where the intra-annual sample variance has quasi-periodic variations and temporal correlations. Such issues lead to inaccurate estimation of parameters coefficients, if they are not handled appropriately. Specifically, the trend detected might be misleading to a large extent. Many studies also suggest that the Long-range
dependence (LRD) in a time series, if a time series displays a slowly declining autocorrelation function (ACF) in variance, can significantly increase the uncertainty of trend detection \citep{yue2002influence,Fatichi2009,Franzke2010,Franzke2012,gao2017quantile}.  Note that LRD can be also reflected from the ACF of residuals in a mean regression model.  To include all the heterogeneity of variance into the model, GARCH model is used to model the variance of temperature time series. \citet{maheu2005can} suggests that GARCH model can capture the shape of the ACF of volatility in daily financial return series, and is consistent with long-memory based on semiparametric and parametric estimates. Hence, our variance model not only accounts for seasonality and temporal correlation, but also reduces the uncertainty caused by LRD.

Spatio-temporal quantile regression has been applied to analyze the temperature series from 72 meteorological stations in Australia. By including the variance term as co-variate in the quantle model, the proposed quantile regression approach has considered all heterogeneity including quasi-periodic variations and temporal correlations. We only include the linear trend in the modeling as both linear and quadratic trend showed a similar pattern during the study period. Hence, our model can detect a reliable trend in the temperature time series.

With analysis of results, we confirm the different patterns of climate change for different percentiles of daily maximum and minimum temperature series over Australia. Overall, the country appears to be experiencing a warming in daily maximum temperature except for a small area in far north QLD, and both cold and hot days are tending to get get warmer. A warming trend in hot days will lead to more frequent extreme heat events. Generally, NSW, south QLD and TAS experience the most significant warming in daily maximum temperature compared with other regions.

In terms of daily minimum temperature,  South QLD and north coast of WA experience more warming than other part of Australia. However, the warming trend is less significant than for daily maximum temperature. The daily minimum temperature series is associated with extreme cold events. Specifically, VIC, TAS and south WA have increasing number of extreme cold days. 
Notably, South QLD experiences an overall climate warming in both maximum and minimum temperature, leading to more frequent hot days and fewer cold days.
 The year round warming trends in South QLD may have significant adverse impacts on agricultural production in Queensland currently aiming for boosting 
 from AUD\$60  to 100 billions by 2030.
  NSW experiences a climate warming in daily maximum temperature with more hot days, but stable daily minimum temperature. VIC and TAS get a more extreme weather, where VIC get much more extreme cold days and slightly more extreme hot days, while TAS has much more extreme hot days and slightly more extreme cold days.  When we look into summer, NSW, VIC, SA and south QLD experience a more warming summer than other regions that have no significant warming trend in summer.  Winter generally gets warmer in most regions in Australia but gets less warmer in VIC with regards to daily maximum temperature. In daily minimum temperature series, it gets quite a bit warmer in QLD, while other regions only get slightly warmer or even colder in south WA, NSW and VIC.

It is worthy nothing that our model results in different changing rates with that presented in regional reports  \citep{Ekst2015,Moise2015,Timbal2015,Watterson2015,Grose2015,Hope2015,McInnes2015}. This is because these reports focused on the period between 1910 and 2013 with using a linear trend to obtain the changing rate. However, as noted in report \citet{BOMCSIRO2018}, most warming in Australia occurs after 1950 and post-1950 period has faster warming rate than pre-1950 period.

There are some limitations we need to take into account when interpreting the results and analysis.
%There are not enough meteorology station are included in the study due to the data quality and study period.
The included stations are not distributed evenly as there are very few stations in the inland region of WA, NT and SA. This could lead to over-generalization of conclusions in these areas. Also, in spatio-temporal quantile regression, we only consider the location coordinates of stations, but we do not to account for the impact of geography, e.g., river catchment, mountain range and distance to ocean. %This might lead some of variation unexplained and incorrect trend %estimate.

\bibliographystyle{ametsoc2014}
% \bibliography{references}
%\bibliographystyle{plain}

\bibliography{refs}

\appendix

\section{Exploratory Analysis}
\label{sec:preli}

\begin{figure}[H]
	\centering
	\includegraphics[width=0.8\textwidth]{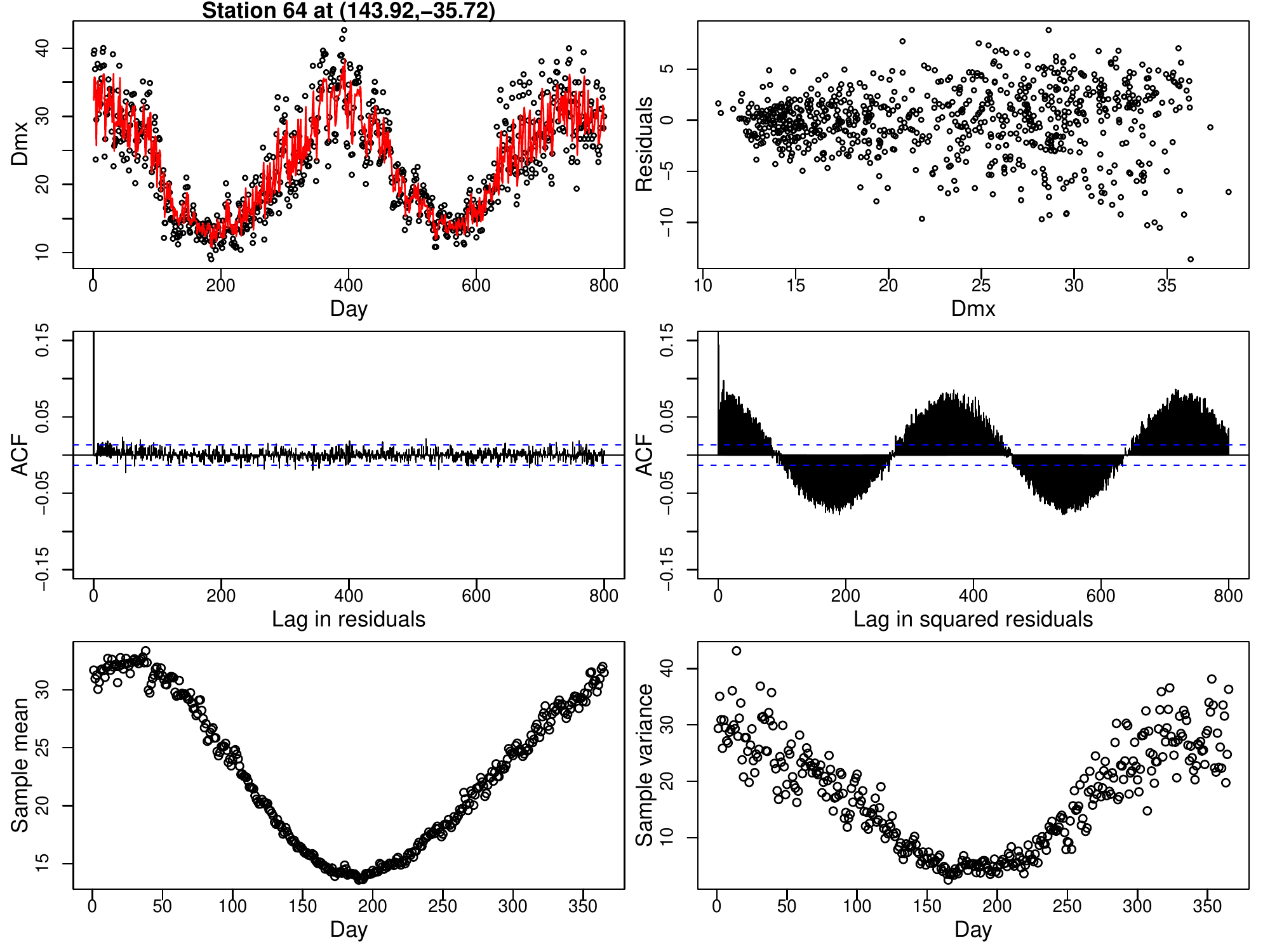}
	\caption{Exploratory analysis for a single station. Top left: fitting parametric mean model for Dmx. Top right: residuals of fit against predicted Dmx for the parametric mean model. Middle left: Auto-Correlation Function (ACF) plot for residuals for lag up to 800 days. Middle right: ACF plot for squared residuals for lag up to 800 days. Bottom left: sample mean over 60 years. Bottom right: sample variance over 60 years. }
	\label{fig.examplefit}
\end{figure}

The daily temperature time series are affected by seasonality, with quasi-periodic variations in both sample mean and variance; see \citet{campbell2005weather,benth2007volatility,benth2007spatial,sirangelo2017stochastic}. An example is shown in the bottom panels in Fig~\ref{fig.examplefit}, where the sample mean and variance of a particular station are calculated over 60 years of observations. Clearly both sample mean and variance have quasi-periodic variations. Other stations {had different patterns, depending on the respective ecosystem, but they} similarly presented this phenomenon.

Such mean periodicity in the series can be easily handled with parametric harmonic functions (or truncated Fourier series). Without considering the seasonality or heterogeneity in variance for the time being, the daily temperature time series can be modelled as follows.
\begin{equation} \label{eq.meanmdl}
y_{t}(\bs) = \mu_{t}(\bs) + \sigma(\bs)\epsilon_t(\bs), \quad \epsilon_t(\bs) \sim_{iid} N(0,1),
\end{equation}
where $t$ is $t$-th day from January 1, 1960 to December 31, 2019, $\bs$ is a particular station, and true value, \[\mu_{t}(\bs)= \beta_{0}(\bs) + \beta_{1}(\bs) t + \mathbf{\beta}_2(\bs) \bx_t(\bs) + FSk(t,\boldsymbol{a}(\bs),\boldsymbol{b}(\bs)) + \sum_{i=1}^p \rho_i(\bs) e_{t-i}(\bs) .\]
{Note that $\epsilon_t(\bs)$ is assumed white-noise and small-scaled and $\sigma^2(\bs)$ is constant in location $\bs$.}
Here $FSk(t,\boldsymbol{a}(\bs),\boldsymbol{b}(\bs))$ is the k-th order truncated Fourier series, and $\sum_{i=1}^p \rho_i(\bs) e_{t-i}(\bs)$ is an $AR(p)$ process.The term $\mathbf{\beta}_2(\bs) \bx_t(\bs)$ is the term for other covariates, e.g., Southern Oscillation Index (SOI) values. For simplicity of notation, we denote the $k$th order truncated Fourier series as follows
\[ FSk(t,\boldsymbol{a}(\bs),\boldsymbol{b}(\bs)) = \sum_{j=1}^{k}\left[ a_{j}(\bs) \sin\left(\frac{2\pi j t}{365}\right) +b_{j}(\bs) \cos\left(\frac{2\pi j t}{365}\right) \right],\]
{where $\boldsymbol{a}(\bs) = (a_1(\bs),\ldots,a_k(\bs))$ and $\boldsymbol{b}(\bs) = (b_1(\bs),\ldots,b_k(\bs))$ are the coefficients of harmonic terms for station $\bs$.}

The red curve in the top left panel of Fig~\ref{fig.examplefit} shows the fitted values of $\mu_{t}(\bs)$ for one station for illustration, these are obtained from model~\ref{eq.meanmdl}. The top right panel shows the residuals against fitted values which clearly indicates the heterogeneity issue in variance. This leads to the failure of the equal variance assumption in model \ref{eq.meanmdl}. Tests of of temporal auto-correlation are shown in the middle panels: for this station, no significant auto-correlation is found for the residuals, while significant and quasi-periodic auto-correlation is found in the squared residuals.

Although the heterogeneity issue is not emphasized or even ignored in many climate themed articles, some models have been proposed to address it. For instance, in the model of the daily average temperature of four US cites by \citet{campbell2005weather}, variance is accounted for via a GARCH process.% as follows
%\begin{equation}
%\sigma_t^2(\bs) = FSk(t,\bs) +  \sum_{r=1}^R\alpha_r(\bs)(\sigma_{t-r}(\bs)\epsilon_{t-r}(\bs))^2 + \sum_{q=1}^{Q} %\beta_q(\bs) \sigma_{t-q}^2(\bs),
%\end{equation}
%where $FS$ denotes a Fourier series. In this variance model, the Fourier series part adequately captures all seasonality %whereas the GARCH part captures the remaining nonseasonal volatility present. It is also concluded that the seasonal %component is relatively more important, and the nonseasonal GARCH part has a smaller effect.
In \citet{benth2005stochastic,benth2007spatial,benth2007volatility}, for example, ARCH models were employed, while  \citet{sirangelo2017stochastic} models the inter-annual sample variance $\sigma_d^2(\bs)$ as simple periodic functions.

%The inter-annual sample variance $\sigma_d^2(\bs)$ was modeled directly in \citet{sirangelo2017stochastic}, where
%\begin{equation}
%\sigma_d^2(\bs) = \beta_0(\bs) + FSk(d,\bs),
%\end{equation}
%where $d = 1,\ldots,365$.

\bigskip
In this study, we model the variance of daily temperature with the sample variance. Let $y_{i, d}(\bs)$ be the maximum (or minimum) daily temperature on the $d$-th day of year $i$ recorded at station $\bs$. Note that we assume there are 365 days for each year and February the 29$^{th}$ was omitted in the analysis.

For each particular day, $d$, for station $\bs$, we compute the inter-annual sample mean and variance, denoted as $\hat{\mu}_{d}(\bs)$ and $\hat{\sigma}_{d}^2(\bs)$:
\[\hat{\mu}_{d}(\bs) = \frac{1}{T}\sum_{i=1}^T y_{i,d}(\bs),\qquad \hat{\sigma}^2_{d}(\bs) = \frac{\sum_{i=1}^T (y_{i,d}(\bs)-\hat{\mu}_{d}(\bs))^2}{T-1}, \]
where $T$ is number of years that are included in the analysis.

The mean is modeled as follows
\begin{equation}
\mu_{d}(\bs) = \hat{\mu}_{d}(\bs) + \epsilon_{d}(\bs),
\end{equation}
where $\epsilon_{d}(\bs) = \hat{\sigma}_{d}(\bs) \delta_{d}(\bs)$ and $\delta_{d}(\bs)\sim_{iid} N(0,1)$.  We include the mean and  squared mean as a covariate of variance function {at the end of the equation below}.

\begin{equation} \label{eq.model4}
\log\left(\sigma^2_{d}(\bs)\right) =  \beta_0(\bs) + \beta_1(\bs)\mu_{d}(\bs) +\beta_2(\bs) \mu_{d}^2(\bs) +FS4(d,\boldsymbol{a}(\bs),\boldsymbol{b}(\bs)) + \rho_1(\bs) \left(\hat{\sigma}_{d-1}^2(\bs) - \sigma^2_{d-1}(\bs)\right) .
\end{equation}

%and model ~\eqref{eq.model4} outperforms other models in terms of AIC %and BIC over all the 72 stations.  The fitted values of the variance function will be used later for spatialtemporal %quantile regression modeling.

We have tested and compared a number of variance models to account for the
inter-annual  heterogeneity that varies from site to site. Different orders of Fourier series have been  tested and selected as $k=4$.  Quantile regression requires the heterogeneity function at each quantile level (except $\tau=0.50$),
we therefore need to jointly estimate the regression parameters and variance parameters.

\section{Joint Models for Quantile Regression and Variability}

Quantile regression permits simultaneous analysis of several features of the response distribution.   To jointly model all quantiles simultaneously and spatially, while accounting for heterogeneity, we propose here an improved version of the spatio-temporal quantile regression approach for Australian daily temperature data, according to \citet{Brian2013}.
\subsection{Spatial-temporal quantile regression with heterogeneity}

Now considering heterogeneity in variance, the model in Eq~\eqref{eq.meanmdl} can be modified as follows:
\begin{equation} \label{eq.qrt1}
y_{t}(\bs) = \mu_{t}(\bs) + \sigma_t(\bs)\epsilon_t(\bs), \quad \epsilon_t(\bs) \sim_{iid} f(\bs),
\end{equation}
where $f(\bs)$ is the PDF of $\epsilon_t(\bs)$ and we denote $F(\bs)\in [0,1]$ be the corresponding CDF at location $\bs$.

The {\it  quantile function} $q(\tau |\bs,t)$ is the function that satisfies $\mathbb{P}\{y_t(\bs)<q(\tau | \bs, t)\}= \tau \in [0,1]$. Inserting Eq~\eqref{eq.qrt1} into this expression we have $\mathbb{P}\{\mu_{t}(\bs) + \sigma_t(\bs)\epsilon_t(\bs)<q(\tau | \bs, t)\}= \tau$ and the quantile function $q(\tau |\bs,t)$ can be expressed as follows.

\begin{equation}\label{eq.qrt2}
q(\tau |\bs,t) = \mu_{t}(\bs)+\sigma_t(\bs)F^{-1}(\tau),
\end{equation}
where $F^{-1}(\tau)$ is the inverse CDF. If the error term $\epsilon_t(\bs)$ is assumed normally distributed, then $F^{-1}(\tau) = \Phi^{-1}(\tau)$. In this study, we are interested in testing the changes in the quantile function $q(\tau |\bs,t)$ over time $t$ for each $\tau$. Therefore, in model Eq~\eqref{eq.qrt2}, the term  $\sigma_t(\bs)$ needs to account for the time variable $t$. For example, for the case of Gaussian distributed response with linear trend over time in \cite{Chandler2005},   the mean  is modeled $\mu_{t}(\bs) = \beta_0(\bs)+\beta_1(\bs)t$, standard deviation $\sigma_t(\bs) = \theta_0(\bs) + \theta_1(\bs) t$ with $F^{-1}(\tau) = \Phi^{-1}(\tau)$. Here, the standard deviation linearly changes with time $t$.
The Gaussian assumption is generalized in \citet{Brian2013}, with incorporating a piece-wise Gaussian basis function so as to describe the linearly changing heterogeneity more flexibly for non-Gaussian distributed response. Furthermore, it is able to characterize the entire quantile process with accommodation of non-Gaussiantiy, such as asymmetry and skewness. However, the model of \citet{Brian2013} was  proposed for a set of annual monthly temperature data without consideration of seasonality in mean and variance. Therefore, it is not applicable to daily temperature data.

In daily temperature data, the quantile function should change periodically according to the season. To model such seasonality, as discussed in our exploratory analysis, we include the $k$-th order truncated Fourier series in the quantile function. Moreover, to investigate the impact of extreme climate events, the Southern Oscillation Index (SOI) is included. The short term changes in Australia's climate are mostly associated with  El Niño or La Niña events that indexed by the SOI.  Including SOI index in the model can remove the short-term natural variability from the long-term warming trend.

Next, we will describe our generalized model which is suitable for daily temperature data. It is worth noting that the model is specifically for trend detection so that the time period $t$ is confined within $[0,1]$ to satisfy the constraints of a quantile function $q(\tau|\bs,t)$ increasing in $\tau$ as suggested in \citet{Brian2013}.  A direct model \eqref{eq.qtstfs4} is as follows.

Let $0=\kappa_1<\ldots<\kappa_{L+1}=1$ be a grid of equally spaced knots covering $[0,1]$. Then, for $l$  with $\kappa_l<0.5$,
\[
B_l(\tau)=\begin{cases}
\Phi^{-1}(\kappa_l)-\Phi^{-1}(\kappa_{l+1}) & \text{if}\quad \tau < \kappa_l \\
\Phi^{-1}(\tau)-\Phi^{-1}(\kappa_{l+1}) &\text{if}\quad \kappa_l \leq \tau < \kappa_{l+1} \\
0 &\text{if}\quad  \kappa_{l+1} \leq \tau
\end{cases}\]
and, for $l$ such that $\kappa_l\geq 0.5$,

\[
B_l(\tau)=\begin{cases}
0 &\text{if}\quad \tau < \kappa_l \\
\Phi^{-1}(\tau)-\Phi^{-1}(\kappa_{l+1}) &\text{if}\quad \kappa_l \leq \tau < \kappa_{l+1} \\
\Phi^{-1}(\kappa_{l+1})-\Phi^{-1}(\kappa_{l}) &\text{if}\quad  \kappa_{l+1} \leq \tau
\end{cases}.\]

And each quantile is assumed to be a function of time $t$ for each station as follows:

\begin{equation}\label{eq.qtstfs4}
\qtst=g_0(\tau| \bs)+ g_1(\tau|\bs)t + g_2(\tau|\bs) x_{soi} + FSk(t,\boldsymbol{g}_s(\tau|\bs),\boldsymbol{g}_c(\tau|\bs)),
\end{equation}
where $\boldsymbol{g}_s(\tau|\bs) = (g_3(\tau|\bs),\ldots,g_{k+2}(\tau|\bs))$, $\boldsymbol{g}_c(\tau|\bs) = (g_{k+3}(\tau|\bs),\ldots,g_{2k+2}(\tau|\bs))$, and for each $k$, $g_k(\tau|\bs)$ is taken to be linear combinations of $L$ basis functions,
\begin{equation}
g_k(\tau|\bs) = \beta_k(\bs) +\sum_{l=1}^L B_l(\tau)\theta_{k,l}(\bs).
\end{equation}

Here $g_1(\tau|\bs)$ is called as \textit{trend function} and is of our particular interest for trend detection in this study.  Note that when $g_k(\tau|\bs) = 0$ for all $k>1$, model Eq \eqref{eq.qtstfs4} degenerates into a linear model that is exactly the same as that used in \citet{Brian2013}.  If further, we let $L=1$ and $B_1(\bs)=\Phi^{-1}(\tau)$,  it is the special case (Gaussian) that corresponds to the model in \cite{Chandler2005}.

And Eq \eqref{eq.qtstfs4} can be further written out,
\begin{equation}\label{eq.qtstfs4_beta_theta}
\begin{split}
\qtst=&\underbrace{\beta_0( \bs)+ \beta_1(\bs)t + \beta_2(\bs) x_{soi} + FSk(t,\boldsymbol{\beta}_s(\bs),\boldsymbol{\beta}_c(\bs))}_{\mu_t(\bs)}   \\
&+\sum_{l=1}^{L}B_l(\tau) \underbrace{\left[\theta_{0,l}(\bs)+\theta_{1,l}(\bs)t + \theta_{2,l}(\bs)x_{soi} +FSk(t,\boldsymbol{\theta}_{s,l}(\bs),\boldsymbol{\theta}_{c,l}(\bs))  \right]}_{\sigma_l(\bs,t)},
\end{split}
\end{equation}
where $\boldsymbol{\beta}_s(\bs)= (\beta_i(\bs),\ldots,\beta_{k+2}(\bs))$, $\boldsymbol{\beta}_c(\bs)= (\beta_{k+3}(\bs),\ldots,\beta_{2k+2}(\bs))$, $\boldsymbol{\theta}_{s,l}(\bs)= (\theta_{3, l},\ldots,\theta_{k+2, l})$, and $\boldsymbol{\theta}_{c,l}(\bs)= (\theta_{k+3, l},\ldots,\theta_{2k+2, l})$. In model Eq \eqref{eq.qtstfs4_beta_theta}, $\beta_k(\bs)$, the center of the quantile function at location $\bs$, and $\theta_{k,l}(\bs)$ are unknown coefficients that determine the shape the quantile function.

Here, adding these terms facilitates the model to account for seasonal heterogeneity in variance. However, auto correlations in both mean and variance are not considered. Moreover, the variance function also depends on the mean values according to the exploratory analysis in Section \ref{sec:preli}. To account for all information, we further improve the model by replacing the  $\theta_{0,l}(\bs) +FSk(t,\boldsymbol{\theta}_{s,l}(\bs),\boldsymbol{\theta}_{c,l}(\bs))$ with $\theta_{2k+3,l}(\bs)\sigma_{d(t)}(\bs)$ in $\sigma_l(\bs,t)$, where $d(t)$ is the $d(t)$-th day of a year for the $t$-th time point and $\sigma_{d(t)}(\bs)$ is modeled in Eq \eqref{eq.model4}. The new model is given as follows
\begin{equation}\label{eq.qtstfs4_beta_theta_sigma}
\begin{split}
\qtst=&\underbrace{\beta_0( \bs)+ \beta_1(\bs)t + \beta_2(\bs) x_{soi} + FSk(t,\boldsymbol{\beta}_s(\bs),\boldsymbol{\beta}_c(\bs))}_{\mu_t(\bs)}   \\
&+\sum_{l=1}^{L}B_l(\tau) \underbrace{\left[\theta_{1,l}(\bs)t + \theta_{2,l}(\bs)x_{soi} + \theta_{2k+3,l}(\bs)\sigma_{d(t)}(\bs) \right]}_{\sigma_l(\bs,t)}.
\end{split}
\end{equation}
Note that the model Eq \eqref{eq.qtstfs4_beta_theta_sigma} is equivalent to adding a new term $g_{2k+3l}(\tau|\bs)\sigma_{d(t)}$ to Eq \eqref{eq.qtstfs4_beta_theta} and turning some unknown parameters ($\beta_{2k+3,l}(\bs)$, $\theta_{0,l}(\bs)$, $\boldsymbol{\theta}_{s,l}(\bs)$ and $\boldsymbol{\theta}_{c,l}(\bs)$) to zeros. Compared with model Eq \eqref{eq.qtstfs4_beta_theta}, model Eq \eqref{eq.qtstfs4_beta_theta_sigma} has far fewer unknown parameters to estimate.  In both spatio-temporal quantile models Eq \eqref{eq.qtstfs4_beta_theta} and \eqref{eq.qtstfs4_beta_theta_sigma}, the quantile function can vary spatially by allowing both the $\beta_k(\bs)$ and $\theta_{k,l}(\bs)$ to be Gaussian spatial processes with exponential covariance. The $\beta_k$ are independent Gaussian processes with mean $\bar{\beta}_k$ and covariance $COV(\beta_k(\bs),\beta_k(\bs'))= \psi_{\beta_k}^2\exp(-||\bs-\bs'||/\rho_{\beta_k})$. The $\theta_{k,l}$ are modeled similarly with mean $\bar{\theta}_{k,l}$ and covariance $COV(\theta_{k,l}(\bs),\theta_{k,l}(\bs'))= \psi_{\theta_k}^2\exp(-||\bs-\bs'||/\rho_{\theta_k})$, but they must satisfy $\sigma_l(\bs,t)>0$ for all $l$ and $t$.

\bigskip
The density function of $y_t(\bs)$ can be expressed in a closed form. Firstly, the quantile function can be written as
\begin{equation}\label{eq.qtst10}
q(\tau|\bs,t) = \sum_{l=1}^{L} \left[a_l(\bs,t) + \sigma_{l}(\bs,t)\Phi^{-1}(\tau) \right] I_{\{\kappa_{l}\leq \tau < \kappa_{l+1}\}},
\end{equation}
where $\sigma_{l}(\bs,t)$ is the coefficient of $B_l(\tau)$ and $a_l(\bs,t) = q(\kappa_{l+1}|\bs,t) - \sigma_{l}(\bs,t)\Phi^{-1}(\kappa_{l+1})$ if $\kappa_{l}<0.5$ and $a_l(\bs,t) = q(\kappa_{l}|\bs,t) - \sigma_{l}(\bs,t)\Phi^{-1}(\kappa_{l})$ if $\kappa_{l} \geq 0.5$. Then the density function of $y_t(\bs)$ is given as follows
\begin{equation} \label{eq.density}
\begin{split}
f(y_t(\bs)) =& \sum_{l=1}^{L} I_{\{q(\kappa_l|\bs,t)< y_t(\bs)\leq q(\kappa_{l+1}|\bs,t)\}} N(y_t(\bs) | a_l(\bs,t),\sigma_l(\bs,t)^2),\\
\end{split}
\end{equation}
%& \sum_{l=1}^{L} I_{\{q(\kappa_l|\bs,t)< y_t(\bs)\leq q(\kappa_{l+1}|\bs,t)\}} N(y_t(\bs) | a_l(\bs,t),\sigma_l(\bs,t)^2)
where $\sigma_l(\bs,t) >0$ for all $l$ and $t$ at each location $\bs$.

\bigskip
It is also worth noting that additional residual correlation is assumed to be an AR(1) process and handled with a copula approach using a latent residual process that is implemented in \citet{Brian2013}. Let $v_t(\bs)$ be a latent Gaussian process and modeled as follows
\begin{equation}
\begin{split}
v_1(\bs) &= w_1(\bs) \\
v_t(\bs) &= \psi_v v_{t-1}(\bs) + \sqrt{(1-\psi_v^2)}w_t(\bs), \quad\textrm{for}\quad t>1,
\end{split}
\end{equation}
where $|\psi_v|<1$ and $w_t(\bs)$ are independent spatial process with mean 0 and covariance $COV(w_t(\bs), w_t(\bs')) = \exp(-||\bs-\bs' ||/\psi_w)$. Here, $v_t(\bs)\sim N(0,1)$ for each $\bs$ and $t$, and let $u_t(\bs) = \Phi(v_t(\bs))\sim U(0,1)$. Let $\tau = u_t(\bs)$ in Eq \eqref{eq.qtst10}, then we have
\begin{equation}\label{eq.qtst_yt}
y_t(\bs) = q(u_t(\bs)|\bs,t) = \sum_{l=1}^{L} \left[a_l(\bs,t) + \sigma_{l}(\bs,t)u_t(\bs) \right] I_{\{\kappa_{l}\leq u_t(\bs) < \kappa_{l+1}\}},
\end{equation}
and
\begin{equation} \label{eq.density_yt}
\begin{split}
f(y_t(\bs)) =& \sum_{l=1}^{L} I_{\{\kappa_{l}\leq u_t(\bs) < \kappa_{l+1}\}} N(y_t(\bs) | a_l(\bs,t),\sigma_l(\bs,t)^2).\\
\end{split}
\end{equation}

\bigskip
As for the model inference,  parameters are estimated in a Bayesian manner via the Markov chain Monte Carlo method that was implemented in \citet{Brian2013}. Note that $\beta_k$ and $\theta_{k,l}$ are model parameters and randomly sampled from their prior distributions (which are Gaussian spatial process) with Gibbs sampling approach. Parameters (like $\psi_{\beta_k}$, $\rho_{\beta_k}$, $\psi_{\theta_{k}}$ and $\rho_{\theta_k}$) are used to characterize the prior distributions and they are updated by using Metropolis sampling with Gaussian candidate distribution. In the residual correlation modeling, the $w_t$ are independent of time $t$ and sampled similarly with $\beta_k$ and $\theta_{k,l}$. Also, $\psi_v$ and $\psi_w$ are distributional parameters and sampled with Metropolis sampling. The closed form density function in Eq \eqref{eq.density} is used to construct the likelihood function.

\subsection{Simulation study}

Let $L=4$, then $\kappa_1=0, \kappa_2 = 0.25, \kappa_3 = 0.5, \kappa_4 = 0.75, \kappa_5 = 1$. Let $y_t(\bs)$ be the simulated data point of day $t$ and location $\bs$, and
%$y_t(\bs) = u_t(\bs) + \epsilon_t(\bs) \sigma_t(\bs)$, where $\epsilon_t(\bs) \sim N(0,1)$ and $u_t(\bs)$
$y_t(\bs)$ follows a piece-wise normal distribution with pdf $f$ that defined in Eq \eqref{eq.density}. Here, for simplicity,
we first set  $\sigma_l(\bs,t) = \theta_{0,l}(\bs)+\theta_{1,l}(\bs)t +FS4(t,\boldsymbol{\theta}_{s,l}(\bs),\boldsymbol{\theta}_{c,l}(\bs))$.
%and $a_l(\bs,t) = q(\kappa_{l+1}|\bs,t)-\sigma_l(\bs,t)\Phi^{-1}(\kappa_{l+1})$ if $\kappa_{l} < 0.5$ and  $a_l(\bs,t) = q(\kappa_{l}|\bs,t)-\sigma_l(\bs,t)\Phi^{-1}(\kappa_{l})$ if $\kappa_{l} \geq 0.5$.
{Then the generated $\{y_t(\bs)\}$ need to be corrected so that the variance can have the similar feature in real series.} The procedure to generate $y_t(\bs)$ is as shown in Procedure \ref{alg.generatedata}.
\begin{algorithm}[H]
	1. Generate $X_t(\bs)\sim N(0,1)$, $t=1,\ldots,T$.\\
	2. Set $y_{t}^{*}(\bs) = \sum_{l=1}^{4}I_{\{\Phi^{-1}(\kappa_l) \leq X_t(\bs)\leq \Phi^{-1}(\kappa_{l+1})\}}X_t(\bs) \sigma_l(\bs,t)+a_l(\bs,t)$ for all $t$. \\
	%\item Let $y_t(\bs) = u_t(\bs) + \epsilon_t(\bs) \sigma_t(\bs)$
	3. Let \[y_t(\bs) = \frac{y_t^{*}(\bs) - \hat{u}_{d(t)}(\bs)}{\hat{\sigma}_{d(t)}(\bs)} \sigma_t(\bs) + \hat{u}_{d(t)}(\bs),\] where $t$ is $d$th day of a year, and $\hat{u}_{d(t)}(\bs)$ and $\hat{\sigma}^2_{d(t)}(\bs)$ are sample mean and variance of generated $y_{t}^{*}(\bs)$.
	\caption{Generation of simulated data}\label{alg.generatedata}
\end{algorithm}

\medskip
%\red{This is new simulation setting. This still failed}

To simulate data with features that are closer to real temperature series, we select 10 stations and do a pilot run of the quantile regression to get their corresponding quantile function $q(\kappa | \bs, t)$ for $\kappa = 0.25, 0.5$ and $0.75$. Then we generate the temperature series according to these quantile functions using the procedures outlined above. The simulated series have the same heterogeneity as the real data.

With these simulated data, we run the model Eq \eqref{eq.qtstfs4_beta_theta} and \eqref{eq.qtstfs4_beta_theta_sigma} separately. The true and estimated values (over 100 simulations) of trend function  are shown in TABLE \ref{tab.compare}. The column RMSE indicates the root mean squared values of difference between true and estimated values. Total RMSE shows the summation of RMSE over all stations and three quantiles. The column $\frac{RMSE (\text{no } \sigma_t)}{RMSE (\sigma_t)}$ shows the ratio between RMSE of estimated values from model without and with $\sigma_t$. For $\frac{RMSE (\text{no } \sigma_t)}{RMSE (\sigma_t)} >1$ it means that including $\sigma_t$ as a covariate improves the model in terms of RMSE. As shown in Table ~\ref{tab.compare}, for most rows (18 of 30), the model including $\sigma_t$ outperforms the model without $\sigma_t$, but there are 5 rows for which including $\sigma_t$ reduces the performance.
In terms of total squared error, the model with $\sigma_t(\bs)$ as a co-variate outperforms the model without $\sigma_t(\bs)$. Hence, we conclude that model \eqref{eq.qtstfs4_beta_theta_sigma} in general can detect the true trend more accurately.
\renewcommand\arraystretch{1}
\begin{table} \small
	
	\caption{Comparison of results (estimates values of trend function) with and without $\sigma_t$ as a covariate}\label{tab.compare}
	\begin{longtable}{c | c | c |c c | c c | c }
		
		Station	& $\tau$		&	True trend	&	Trend (no $\sigma_t$)	&	RMSE	&	Trend ($\sigma_t$)	& RMSE	&	$\frac{RMSE (\text{no } \sigma_t)}{RMSE (\sigma_t)}$		\\ \hline
		1	&	tau=0.25	&	0.43	&	0.51	&	0.09	&	0.53	&	0.11	&	0.86	\\
		&	tau=0.5	&	0.59	&	0.58	&	0.06	&	0.6	&	0.03	&	1.75	\\
		&	tau=0.75	&	0.49	&	0.55	&	0.08	&	0.57	&	0.08	&	0.96	\\ \hline
		2	&	tau=0.25	&	1.03	&	0.91	&	0.15	&	0.97	&	0.07	&	2.14	\\
		&	tau=0.5	&	0.80	&	0.73	&	0.11	&	0.76	&	0.06	&	1.69	\\
		&	tau=0.75	&	0.49	&	0.57	&	0.10	&	0.61	&	0.12	&	0.83	\\ \hline
		3	&	tau=0.25	&	1.12	&	1.19	&	0.14	&	1.17	&	0.08	&	1.76	\\
		&	tau=0.5	&	0.91	&	1.02	&	0.16	&	0.97	&	0.10	&	1.68	\\
		&	tau=0.75	&	0.68	&	0.81	&	0.15	&	0.82	&	0.14	&	1.04	\\ \hline
		4	&	tau=0.25	&	0.14	&	0.14	&	0.02	&	0.19	&	0.05	&$\mathbf{0.37}$	\\
		&	tau=0.5	&	0.22	&	0.26	&	0.06	&	0.29	&	0.07	&	$\mathbf{0.78}$	\\
		&	tau=0.75	&	0.26	&	0.27	&	0.03	&	0.31	&	0.05	&	$\mathbf{0.63}$	\\ \hline
		5	&	tau=0.25	&	1.01	&	1.11	&	0.14	&	1.16	&	0.15	&	0.96	\\
		&	tau=0.5	&	0.77	&	0.83	&	0.09	&	0.83	&	0.07	&	1.38	\\
		&	tau=0.75	&	0.96	&	0.92	&	0.09	&	0.93	&	0.03	&	2.86	\\ \hline
		6	&	tau=0.25	&	1.05	&	0.98	&	0.11	&	1.05	&	0.02	&	4.97	\\
		&	tau=0.5	&	1.94	&	1.29	&	0.66	&	1.4	&	0.55	&	1.21	\\
		&	tau=0.75	&	1.12	&	0.85	&	0.28	&	0.9	&	0.22	&	1.29	\\ \hline
		7	&	tau=0.25	&	-1.90	&	-1.75	&	0.21	&	-1.87	&	0.05	&	4.61	\\
		&	tau=0.5	&	-1.42	&	-1.31	&	0.16	&	-1.3	&	0.13	&	1.25	\\
		&	tau=0.75	&	-0.86	&	-0.95	&	0.12	&	-0.93	&	0.08	&	1.52	\\ \hline
		8	&	tau=0.25	&	-0.47	&	-0.54	&	0.09	&	-0.48	&	0.04	&	2.13	\\
		&	tau=0.5	&	0.46	&	0.22	&	0.24	&	0.38	&	0.10	&	2.53	\\
		&	tau=0.75	&	-0.03	&	-0.13	&	0.11	&	-0.07	&	0.06	&	1.95	\\ \hline
		9	&	tau=0.25	&	-0.05	&	0.08	&	0.15	&	0.19	&	0.27	&$\mathbf{	0.56}$	\\
		&	tau=0.5	&	0.40	&	0.64	&	0.27	&	0.66	&	0.30	&	0.91	\\
		&	tau=0.75	&	0.65	&	0.83	&	0.20	&	0.78	&	0.14	&	1.44	\\ \hline
		10	&	tau=0.25	&	0.14	&	0.21	&	0.10	&	0.26	&	0.15	&$\mathbf{	0.64}$	\\
		&	tau=0.5	&	0.07	&	0.38	&	0.35	&	0.38	&	0.36	&	0.99	\\
		&	tau=0.75	&	0.32	&	0.46	&	0.18	&	0.48	&	0.20	&	0.90	\\ \hline
		Total RMSE&		&		&		&	4.71	&		&	3.86	&		\\
	\end{longtable}
\end{table}

\iffalse
\red{The following is the old setting. It failed to show the improvement of using $\sigma_t$}

The $q(\kappa_l|\bs,t)$ is computed with using $\beta_k(\bs)$ and $\theta_{kl}(\bs)$ which are randomly generated from Gaussian spatial process. The $\beta_k$ are assumed to have zero mean and covariance $cov(\beta_k(\bs),\beta_k(\bs'))=v_k^2\exp(-||\bs-\bs'||/\rho_k)$. The $\theta_{kl}$ have covariance $cov(\theta_{kl}(\bs),\theta_{kl}(\bs'))=w_{k}^2\exp(-||\bs-\bs'||/\phi_k)$ and satisfy the monotonicity (increasing) of quantiles, where $\sum_{k=1}^{p}\theta_{kl}(\bs)x_k(\bs) >0$.

For simplicity of simulation, I only include two variables intercept and $t$, so that each location have 10 unknown parameters.  In this setting the quantile function $q(\tau | \bs, t)$ is a linear function of time.
\begin{enumerate}
	\item Set $v_k = 1$ and $\rho_k=1$, then compute covariance and generate $\beta_k$ from multivariate normal distribution for $k=1,2$.
	\item Set $w_k =1 $ and $\phi_k = 1$, then compute covariance and generate $\theta_{kl}$ from multivariate normal distribution for $k=1,2$. Check and accept if $\theta_{0l}+\theta_{11}>0$.
\end{enumerate}
\fi

\end{document}